# Parker Solar Probe Observations of Compound Reconnection Exhaust Boundaries and Mirror-Mode Structures in the Near-Sun Heliospheric Current Sheet

Weijie Sun[1] (weijiesun@berkeley.edu), Tai Phan[1], Jia Huang[1], Yi-Hsin Liu[2], James A. Slavin[3], Orlando Romeo[1], Mingzhe Liu[1], Vassilis Angelopoulos[4], Ali Rahmati[1], Davin Larson[1], Nehpreet Walia[5], Stuart Bale[1], Marc Pulupa[1], Jiutong Zhao[1], Roberto Livi[1]

1, Space Sciences Laboratory, University of California, Berkeley, CA, USA
2, Department of Physics and Astronomy, Dartmouth College, Hanover, NH, USA
3, Department of Climate and Space Sciences and Engineering, University of Michigan, Ann Arbor, MI, USA
4, Earth, Planetary and Space Sciences, and Institute of Geophysics and Space Physics, University of California, Los Angeles, CA, USA
5, Department of Astrophysical Sciences, Princeton University, NJ 08544, USA

## Abstract

Magnetic reconnection is a fundamental physical process that can drive rapid conversion of magnetic energy into plasma bulk flows, thermal heating, and particle acceleration in space and astrophysical plasmas. Classical reconnection theory predicts that the Alfvénic reconnection exhausts are bounded by pairs of slow-mode shocks. However, identifying and characterizing these shocks through *in situ* spacecraft observations remains a challenge. Here we report Parker Solar Probe (PSP) observations of a reconnection exhaust embedded in the heliospheric current sheet (HCS) at a heliocentric distance of 12.2 $R_\odot$. The reconnection exhaust is bounded on both boundaries by compound magnetic structures rather than a pair of pure slow shocks. Each boundary consists of a rapidly evolving, steep inner slow shock, whose Mach numbers and shock-normal angles change significantly within several minutes, and an outer, gradual compound structure which comprises a slow shock and a rotational discontinuity. These slow shocks are quasi-perpendicular and are accompanied by enhanced proton perpendicular heating. Deep within the reconnection exhaust, high perpendicular temperature together with large plasma $\beta$ trigger mirror instability and generate mirror-mode structures. These observations provide new insights into the structure of reconnection exhaust boundaries and their role in energy conversion in the near-Sun plasma.

## 1. Introduction

Magnetic reconnection is a fundamental plasma process that can effectively convert magnetic energy into particle kinetic and thermal energy. It was originally proposed to explain the fast energy release in solar flares (Parker 1957; Sweet 1958), and since then it has been recognized as a key physical process in heliospheric and astrophysical plasmas. In classical reconnection theory, a localized magnetic diffusion region is centered within an X-type magnetic field configuration, where pairs of slow-mode shocks bound the Alfvénic reconnection exhausts on the two sides (Petschek 1964; see also, Sonnerup 1970; Yeh & Axford 1970). These slow shocks are proposed to accelerate upstream plasma into the reconnection exhausts and promote magnetic energy conversion. However, subsequent theoretical and simulation studies have shown that slow shocks are often replaced (Levy, Petschek, & Siscoe 1964) or accompanied by rotational discontinuity but are well separated from each other, as demonstrated in magnetohydrodynamic (MHD) (e.g., Hoshino & Nishida 1983; Heyn et al. 1988; Lin & Lee 1993) and hybrid (e.g., Lin & Lee 1993; Lin & Swift 1996; Walia, Seki, & Amano 2022). Furthermore, the slow shock and rotational discontinuity may merge to form compound magnetic structures, as demonstrated in hybrid (e.g., Lee et al., 2000; Weng et al., 2012) and particle-in-cell simulations (e.g., Liu, Drake, & Swisdak 2011a, 2011b, 2012; Innocenti et al. 2015).

The magnetic boundary structures of reconnection exhausts have been analyzed based on the *in situ* measurements in several space plasma environments. At Earth's dayside magnetopause under asymmetric conditions, both rotational discontinuities (e.g., Sonnerup & Cahill Jr. 1967; Sonnerup et al. 1981; Paschmann et al. 1986) and slow shocks (Sonnerup et al. 2016; Walia et al. 2018) have been reported in reconnection exhaust boundaries. Biernat et al. (1989) proposed that both rotational discontinuities and slow shocks are needed for the asymmetric magnetopause reconnection with a non-zero guide field. In Earth's magnetotail under more symmetric conditions, slow shocks (Feldman et al. 1984; Smith et al. 1984; Saito et al. 1995; Eriksson et al. 2004; Walia et al. 2024) and compound structures (Whang et al. 1997) have been observed. Similar boundary structures have been also identified in the solar wind current sheets (Gosling et al. 2006; Teh et al. 2009), at the boundaries of large-scale interaction regions (Zuo, Wei, & Feng 2006; Zhou et al. 2018; Duan et al. 2023), and in the heliospheric current sheet (HCS) (Zhou et al. 2022). However, statistical survey shows that the occurrence rate of slow shocks is ~ 20% of the reconnection exhausts at Earth's dayside magnetopause (Walia, et al. 2018) and ~ 40 to 55% in Earth's magnetotail (Walia, et al. 2024).

Despite these theories and observations, the physical properties and the role of slow shocks and compound structures in magnetic reconnection are still not well understood. In this letter, we report Parker Solar Probe (PSP) (Fox et al. 2016) observations of an anti-sunward reconnection exhaust embedded in the HCS at a heliocentric distance of 12.2 solar radii ($R_\odot$). This investigation provides new insights into the magnetic boundaries and the roles of slow shocks in the magnetic reconnection, which further improve our understanding of reconnection generated slow shocks and their role in magnetic energy conversion in the near-Sun plasma.

## 2. Reconnection Exhaust on 30 September 2024

### 2.1. Data Sources and Instrumentations

This study utilizes vector magnetic field measurements at a cadence of ~ 0.22 s from the FIELDS instrument suite (Bale et al. 2016). Plasma measurements are provided by the Solar Wind Electrons Alphas and Protons (SWEAP) investigation (Kasper et al. 2016). Proton moments and energy spectra with a time resolution of ~ 3.5 s are from the Solar Probe ANalyzers-Ion (SPAN-i) (Livi et al. 2022), while electron measurements are from the Solar Probe ANalyzers-Electron (SPAN-e) (Whittlesey et al. 2020). Electron number density is also independently provided by quasi-thermal noise (QTN) spectroscopy based on radio-frequency measurements (Moncuquet et al. 2020) and the spacecraft surface charging method (Probe) using the FIELDS voltage probes (Liu et al. 2025). The "Probe" density used here has a temporal resolution of 0.1 s. In the calculations associated with plasma density, the QTN density is used.

All vector quantities are shown under the RTN coordinate system, where $\hat{R}$ points radially away from the Sun, $\hat{T}$ is directed along solar rotation, and $\hat{N}$ completes the right-handed coordinate system, which is approximately aligned with the solar rotational axis. Therefore, the proton velocity is in the Sun-centered reference frame with spacecraft's velocity accounted for.

### 2.2. Overview of Reconnection Exhaust in the Heliospheric Current Sheet

On 30 September 2024, during its 21st perihelion encounter of the Sun, PSP crossed the HCS and observed an anti-sunward reconnection exhaust in the reference frame of the ambient solar wind (Figure 1). The HCS crossing is identified by a polarity reversal of the radial magnetic field component ($B_R$) from negative to positive (Figure 1b), which is accompanied by a reversal in the suprathermal (strahl) electron pitch angles from 180° to 0° (Figure 1h). Within the HCS, electrons are observed at all pitch angles with reduced fluxes in the parallel and antiparallel pitch angles compared to the external solar wind. This anti-sunward reconnection exhaust persists for approximately 50 minutes from ~ 11:50 to 12:40 UTC.

To investigate the reconnection exhaust, we quantitatively evaluate the Walén relation (Hudson 1971; Paschmann, et al. 1986; Sonnerup, et al. 1981), which predicts the local velocity vectors across a rotational discontinuity:

$$\vec{V}_{predicted} = \overline{\vec{V}_1} \pm \left[ \overline{\vec{B}_2}(1-\alpha_2)^{1/2}(\mu_0 \rho_2)^{-1/2} - \overline{\vec{B}_1}(1-\overline{\alpha_1})^{1/2}(\mu_0 \overline{\rho_1})^{-1/2} \right] \quad (1)$$

where the $\vec{B}$ is magnetic field vector, $\rho$ is the plasma mass density, and $\mu_0$ is the magnetic permeability of free space. The pressure anisotropy factor, $\alpha$, is $\mu_0(p_{para} - p_{perp})/B^2$, where $p_{para}$ and $p_{perp}$ are the parallel and perpendicular thermal pressures. In this equation, subscript "1" denotes the averaged reference values measured in the upstream region, while subscript "2" denotes the time-varying local quantities in the downstream region. The sign ± is chosen based on whether the observed velocity and magnetic field variations across the boundary are correlated or anti-correlated.

When applying the Walén relation, an interval of external solar wind plasma needs to be selected serve as the upstream "1". We selected one preceded the entry of the reconnection exhaust from 11:50:00 to 11:51:50 UTC and another after leaving the reconnection exhaust from 12:38:40 to 12:41:30 UTC. The $\vec{V}_1$, $\vec{B}_1$, $\rho_1$ and $\alpha_1$ were averaged over each of the intervals and then applied to the Walén relation separately to calculate the predicted velocity time series. When calculating $\vec{V}_{predicted}$ from the pre-entry interval, a positive sign was used because the observed velocity and magnetic field vectors across the boundary were correlated. A negative sign was used for the post-exit interval as the velocity and field vectors were anti-correlated. As shown in Figure 1f, the R component of $\vec{V}$ in the HCS near the leading edge was consistent with the R component of the $\vec{V}_{predicted}$ from the pre-entry interval and near the trailing edge was consistent with the prediction from the post-exit interval. These features suggest that the external solar wind serves as the inflow (upstream) region for the reconnection exhaust in the HCS.

### 2.3. Compound Slow Shock and Rorational Discontinuity Structures at Reconnection Exhaust Boundaries

Figure 2 shows measurements of the magnetic boundaries during the intervals that PSP entered the reconnection exhaust (left column) and exited the exhaust (right column). During entering the exhaust, from ~ 11:52:10 to 11:52:30 UTC indicated by the pink bar above Figure 2a, PSP observed a gradual decrease in magnetic field intensity from ~ 1000 nT to ~ 900 nT (Figure 2a), which was accompanied by gradual increases in plasma density and bulk velocity (Figures 2c and 2e). The proton temperature showed a sharp increase near the end of this time interval at ~ 11:52:30 UTC (Figure 2d). These changes in magnetic and plasma parameters are consistent with a slow-mode shock transition from upstream to downstream.

To quantitatively determine the normal direction of this magnetic boundary, we apply two independent techniques: minimization of the Rankine-Hugoniot (R-H) relations (**Appendix A**) and minimization based on Whang's shock theory (**Appendix B**). Both techniques use magnetic field and plasma parameters on two sides of the boundary to determine the best-fit normal directions. The results of the two techniques and their best-fit normals are shown in Figures 3a and 3b, respectively. The R-H minimization provides a normal vector of ±[-0.119, -0.971, 0.208], which differs by ~ 4.7° from the normal obtained using Whang's theory.

Using the best-fit normals, the properties of the boundary are determined (see **Appendix C** for calculation details). Under the R-H normal, the upstream and downstream slow magnetosonic Mach numbers ($M_{SMS}$) are ~ 1.09 and ~0.67, respectively, while both remain sub-Alfvénic, with Alfvén Mach numbers ($M_A$) of ~0.57 upstream and ~ 0.47 downstream. Therefore, this boundary satisfies a slow shock. Moreover, the angles $\left(\theta\langle\vec{n}, \vec{B}\rangle\right)$ between the shock normal and the upstream and downstream magnetic fields decrease from ~85.7° to ~84.7°, which confirms the magnetic field feature of the slow shock and further demonstrates that this slow shock is quasi-perpendicular. Under Whang's shock normal, the upstream and downstream $M_{SMS}$ are ~ 1.28 and ~0.82, $M_A$ are ~0.65 and ~ 0.57, and $\theta\langle\vec{n}, \vec{B}\rangle$ are ~81.6° to ~80.2°, consistent with slow shock.

Furthermore, hodogram analysis of the magnetic field across a shorter interval within the slow shock reveals a rotation of the magnetic field vectors in the plane perpendicular to the shock normal (Figure 3d), which is evidence of the presence of a rotational discontinuity. We note that the magnetic field rotates largely along the guide field direction of the HCS. Thus, these results together show that this outer boundary of the reconnection exhaust consists of a slow shock and a rotational discontinuity, which is therefore a compound structure. Both the proton parallel and perpendicular temperatures are enhanced across this compound structure (Figure 2d).

Following this compound structure, at ~11:52:37 UT (the first vertical magenta line from left in Figure 2), the magnetic field intensity decreases further and sharper from ~900 nT to ~400 nT, while the plasma density, proton perpendicular temperature, and bulk velocity increase further. Applying the same analysis techniques demonstrates that this transition can also be interpreted as a quasi-perpendicular slow shock with $M_{SMS}$ of ~1.39 upstream and ~0.96 downstream, $M_A$ of ~0.97 upstream and ~0.84 downstream, and $\theta\langle\vec{n}, \vec{B}\rangle$ of 85.1° upstream and 81.0° downstream with R-H normal, respectively. Although the normal from Whang's shock theory was ~ 5.2° from the R-H normal, $M_{SMS}$ are ~ 1.19 and ~0.89, $M_A$ are ~0.83 and ~ 0.79, and $\theta\langle\vec{n}, \vec{B}\rangle$ are ~84.9° to ~81.3° are consistent with slow shock as well. PSP subsequently crossed this inner slow shock four additional times. All the transitions satisfy quasi-perpendicular slow shocks as summarized in Table 1. We interpret these multiple crossings as repeated encounters of a wavy quasi-planar slow shock due to similar upstream magnetic field strength, proton temperatures, and ion and electron spectral characteristics.

Since this steep transition was observed deeper within the reconnection exhaust, we term this structure as the inner slow shock and previous gradual one the outer slow shock. The features confirm the inner slow shock is a separate structure from the outer slow shock as their upstream magnetic field intensity (~ 900 nT) was always lower than the upstream magnetic field intensity of the outer slow shock (~ 1000 nT). The multiple crossings indicate that this inner slow shock is rapidly evolving, which is confirmed by the large changes in the shock-normal angles and shock Mach numbers among the multiple crossings within several minutes (Table 1).

When PSP later exited the reconnection exhaust and traversed the other side boundary, a similar layered magnetic structure was observed. The spacecraft first crossed the inner slow shock multiple times before encountering the outer compound structure consisting of a rotational discontinuity and a slow shock. This inner slow shock is also rapidly evolving as shown (Table 1). The rotational discontinuity is located within the outer smoother slow shock but was adjacent to the sharper inner slow shock. In addition, both the inner and outer slow shocks are quasi-perpendicular and each crossing from upstream to downstream is associated with an increase in proton perpendicular temperature.

### 2.4. Shock-Driven Perpendicular Temperature Anisotropy and Mirror-Mode Structures Inside Reconnection Exhaust

Both the outer compound structure and the inner slow shock correspond to perpendicular heating of protons from upstream to downstream. This perpendicular heating is in addition to the parallel heating associated with magnetic reconnection (e.g., Gosling et al. 2005; Egedal, Daughton, & Le 2012; Haggerty et al. 2015; Phan et al. 2022). Although in the exhaust boundaries, the proton parallel temperature generally exceeds the perpendicular temperature due to escaping ions, deeper inside the reconnection exhaust the proton perpendicular temperature can become larger than the parallel temperature.

Figure 4 shows the measurements of the reconnection exhaust within the HCS, during which the magnetic field intensity exhibits clear wave-like structures with depressions and peaks coexisting (Figure 4a). In this wave-like structure, the magnetic field intensity is anticorrelated with the plasma number density (Figures 4e and 4f), while the magnetic field direction remains nearly constant (Figures 4g and 4h). These characteristics are consistent with mirror-mode structures. Meanwhile, inside the HCS, especially when the proton perpendicular temperature exceeds the parallel temperature. the mirror-instability threshold is satisfied, i.e., $R = (T_{perp}/T_{para})/(1 + 1/\beta_{perp}) > 1$ (e.g., Hasegawa 1969; Gary, Fuselier, & Anderson 1993) (Figure 4d). This indicates that the observed mirror-mode structures are locally generated by mirror instability. We note that the mirror unstable environment (Huang et al., 2023) and mirror-mode structures (Fargette et al. 2026) are recently reported within the HCS with PSP's measurements.

Furthermore, electron temperature also increases across the slow shocks (Figure 1e), however, the increases are smaller than those of protons. This feature is consistent with the observations of slow shocks in Earth's magnetotail (e.g., Feldman et al. 1987; Saito, et al. 1995) and those shown in numerical study (Daughton et al. 2001).

## 3. Discussion

### 3.1. Compound Magnetic Structures Bounding Reconnection Exhaust

Theoretical studies have predicted several types of compound magnetic structures at the boundaries of reconnection exhausts, including a rotational discontinuity attached the leading edge of a slow shock (RD-SS), a rotational discontinuity attached the trailing edge of a slow shock (SS-RD) and a rotational discontinuity nested within a slow shock (SS-RD-SS) (Lee et al. 2000; Liu, et al. 2011b; Weng et al. 2012). The magnetic boundaries of the reconnection exhaust reported by this study are most consistent with the SS-RD-SS compound structure, in which a rotational discontinuity is nested within the slow shock transition.

The formation of such compound structures has been attributed to the leakage of downstream protons to the upstream near the reconnection exhaust boundary. The leaked protons enhance the parallel ion temperature and produce strong parallel temperature anisotropy, which reduces the intermediate-mode phase speed and thus slows the propagation of the rotational discontinuity relative to the slow shock (e.g., Lee, et al. 2000; Hau & Hung 2005; Liu, et al. 2011b). As a result, the rotational discontinuity would be trapped within the slow shock and form an SS-RD-SS compound structure (Lee, et al. 2000; Weng, et al. 2012). The leakage of downstream protons to the upstream and parallel temperature anisotropy are clearly evident in our event (Figure 1d) and have been studied in other reconnection events within the HCS (Phan et al. 2022, 2025).

The type of magnetic boundary bounding reconnection exhaust is controlled by the magnetic shear angle and upstream ion plasma $\beta$ (Weng, et al. 2012). In the present case, the magnetic shear angle across the HCS is approximately 166° and the plasma $\beta$ is ~ 0.4 upstream of the leading boundary and ~ 0.2 upstream of the trailing boundary. These parameters fall within the regime identified by Weng et al. (2012; their Figure 5) in which SS-RD-SS compound structures are anticipated to form.

These results demonstrate that magnetic reconnection in the HCS or even in the lower solar corona can serve as a source of compound slow shock structures observed in the solar wind (Whang, et al. 1997; Whang et al. 1998). However, the present study only investigates one reconnection event in the HCS. The examination of a larger number of reconnection events, including but not limited to HCS, would enable a more comprehensive categorization of reconnection exhaust boundaries and their role in plasma heating.

### 3.2. Perpendicular Heating and Mirror-Mode Structures

Our observations show that the compound slow shock structures are associated with clear enhancements in proton perpendicular temperature. However, the observed perpendicular heating cannot be explained by simple adiabatic invariance across the slow shocks, as the magnetic field magnitude decreases, the perpendicular pressures for both protons and electrons increase from upstream to downstream. This magnetic pressure decreases of slow shocks contrasts with the magnetic compression of fast-mode shocks. Instead, the perpendicular heating likely reflects non-adiabatic plasma dynamics in slow shock environments and possibly additional heating mechanisms. Previous hybrid simulations have shown perpendicular proton heating downstream of slow shocks, which is attributed to wave-particle interactions and kinetic processes within and downstream of the shock layer (Sato 1979; Lin & Lee 1991; Walia, et al. 2022). Such processes are expected to be particularly effective in the dynamically evolving compound shock structures observed in this study.

The enhanced perpendicular heating across the slow shock drives the plasma within the reconnection exhaust toward conditions favorable for instabilities. The observed temperature anisotropy combined with the large plasma $\beta$ in HCS satisfies the threshold for the mirror instability (e.g., Hasegawa 1969; Gary, Fuselier, & Anderson 1993), which is consistent with the mirror-mode structures observed in the reconnection exhaust. This scenario that slow-shock-driven perpendicular heating followed by the development of mirror instability suggests that magnetic reconnection can actively regulate high-$\beta$ plasma environments and may represent an efficient source of mirror-mode structures in the solar wind (e.g., Winterhalter et al. 1995; Xiao et al. 2010; Yu et al. 2021).

## 4. Conclusions

This study reports PSP observations of a reconnection exhaust in the near-Sun HCS at a heliocentric distance of 12.2 $R_\odot$. The reconnection exhaust boundaries are SS-RD-SS compound magnetic structures, i.e., a rotational discontinuity nested with slow shock transitions, rather than pure slow shocks. The inner slow shock is rapidly evolving and is steeper than the outer slow shock in the changes of magnetic field and plasma parameters. The slow shocks are quasi-perpendicular and are associated with strong proton perpendicular heating, which indicates non-adiabatic and wave-particle processes in energy conversion rather than adiabatic compression. The resulting perpendicular temperature anisotropy, together with large plasma $\beta$ deep in the reconnection exhaust, drives mirror instability and generates mirror-mode structures in the HCS.

The identification of a rotational discontinuity nested with slow shock transitions, which is consistent with theoretical prediction, provides new insight into the reconnection exhaust boundaries. Further statistical investigations will be important in determining how exhaust boundary structures depend on ambient plasma and field conditions. Our results also suggest that the compound slow shock structure can be a driver of plasma acceleration and mirror instability in the near-Sun solar wind.

## Acknowledgement

We are grateful to the Parker Solar Probe team for their data access and support. Parker Solar Probe was designed, built, and is now operated by the Johns Hopkins Applied Physics Laboratory as part of NASA's Living with a Star (LWS) program (contract NNN06AA01C). Support from the LWS management and technical team has played a critical role in the success of the Parker Solar Probe mission. Thanks to the Solar Wind Electrons, Alphas, and Protons (SWEAP) team for providing data (PI: Justin Kasper, BWX Technologies). Thanks to the FIELDS team for providing data (PI: Stuart D. Bale, UC Berkeley). Weijie Sun would like to thank Prof. Zuyin Pu from Peking University, whose insightful lessons and discussions greatly inspired this work. Weijie Sun thanks Zilu Zhou for discussions on Whang's theoretical shock study.

Data used in this study are available at: http://fields.ssl.berkeley.edu/data/ and http://sweap.cfa.harvard.edu/pub/data/sci/sweap

Software: SPEDAS (Angelopoulos et al. 2019).

## Appendix A: Minimization in the Rankine-Hugoniot (R-H) relations

The Rankine-Hugoniot (R-H) relations describe the jump conditions across magnetohydrodynamic (MHD) structures such as shocks and discontinuities in space plasmas (Baumjohann & Treumann 2012). They express the conservation of key plasma quantities across a boundary that separates two distinct plasma regimes.

In the shock rest frame, the R-H relations can be written as follows:

Mass flux ($F_{mass}$) conservation

$$[\rho V_n] = 0 \tag{A1}$$

Normal magnetic field ($B_n$) conservation

$$[B_n] = 0 \tag{A2}$$

Momentum conservation, which includes the normal ($M_n$) and tangential ($M_{tan}$) components

$$[\rho V_n^2 + P + B_{tan}^2/2\mu_0] = 0 \tag{A3a}$$
$$[\rho V_n V_t - B_n B_{tan}/\mu_0] = 0 \tag{A3b}$$

Magnetic flux ($F_{mag}$) conservation, which is equivalent to Faraday's law

$$[V_n B_{tan} - B_n V_{tan}] = 0 \tag{A4}$$

Energy flux ($F_{energy}$) conservation

$$\left[V_n\left(\frac{1}{2}\rho V^2 + \frac{\gamma}{\gamma-1}P + \frac{B_{tot}^2}{\mu_0}\right) - \frac{B_n}{\mu_0}(\vec{V}\cdot\vec{B})\right] = 0 \tag{A5}$$

We need to note that these equations assume ideal MHD, which means that kinetic effects such as finite Larmor radius effects and wave-particle interactions are neglected. In our calculation, temperature anisotropy is also neglected by using only the perpendicular temperature.

To determine the most consistent boundary normal direction, we evaluate for each trial normal on a 180 × 360 grid (1° spacing in both azimuth and elevation) the following equation

$$\sigma_{\theta\phi} = \sqrt{\sum_X \left[(X_1 - X_2)/\left(\frac{X_1}{2} + \frac{X_2}{2}\right)\right]^2} \tag{A6}$$

, in which $X$ includes mass flux $(F_{mass})$, normal magnetic field $(B_n)$, normal momentum $(M_n)$, tangential momentum $(M_{tan})$, magnetic flux $(F_{mag})$, and energy flux $(F_{energy})$.

The angle grid corresponding to the minimum value of $\sigma_{\theta\phi}$ is taken as the best estimate of the boundary normal direction.

## Appendix B: Minimization in Whang's shock theory

In Whang (1987)'s theory, the downstream tangential magnetic field $(B_{tan2})$ can be obtained by solving the cubic equation:

$$B_{tan2}^3 + D_2 B_{tan2}^2 + D_1 B_{tan2} + D_0 = 0 \tag{B1}$$

, where the coefficients $D_i$ are functions of upstream physical quantities.

$$\begin{cases} D_0 = -(\gamma+1)(M_A^2-1)^2 B_n^2 B_{tan1} \\ D_1 = [\mu_0 \gamma P_1 - 2B_n^2 + \gamma B_{tan1}^2 + (\gamma-1)(M_A^2-1)B_n^2](M_A^2-1) \\ D_2 = [1 + (2-\gamma)(M_A^2-1)]B_{tan1} \end{cases} \tag{B2}$$

Here $M_A$ is the upstream Alfvén Mach number,

$$M_A = V_{n1}/(V_{A1} \cos\theta_1) \tag{B3}$$

To obtain $V_{n1}$, both downstream and upstream plasma densities and flow velocities are needed.

Solving equation B1 will provide no more than one shock solution. Once $B_{tan2}$ is determined, the downstream plasma parameters can be derived as follows:

Mass density $(\rho_2)$:

$$\frac{\rho_2}{\rho_1} = M_A^2 B_{tan2} / [(M_A^2-1)B_{tan1} + B_{tan2}] \tag{B4}$$

Downstream normal velocity $(V_{n2})$:

$$V_{n2} = V_{n1}/(\rho_2/\rho_1) \tag{B5}$$

Downstream thermal pressure $(P_2)$

$$P_2/P_1 = 1 - (B_{tan2} - B_{tan1})[B_{tan1} + B_{tan2} - 2(M_A^2-1)B_{n1}^2/B_{tan2}]/(\mu_0 P_1) \tag{B6}$$

From these, the downstream temperature $T_2$ can also be calculated.

To identify the most consistent shock normal, we calculate

$$\delta_{\theta\phi} = \sqrt{\sum_X \left(1 - \frac{X_{whang}}{X_{PSP}}\right)^2} \qquad (B7)$$

, in which $X$ includes mass density ($\rho_2$), temperature ($T_2$), thermal pressure ($P_2$), normal velocity ($V_{n2}$), tangential magnetic field component ($B_{tan2}$), and normal magnetic field ($B_{n2}$).

The angular grid corresponding to the minimum $\delta_{\theta\phi}$ is taken as the best estimate of the shock normal direction. A similar technique may have been applied to determine the shock normal direction by several studies (See, Zuo, et al. 2006; Zhou, et al. 2022).

## Appendix C: Calculations of the Shock Parameters

With the shock normal determined, we then calculated the shock propagation speed using the conservation of momentum flux across the shock:

$$V_{sh,n} = \frac{N_d \vec{V}_d \cdot \vec{n} - N_u \vec{V}_u \cdot \vec{n}}{N_d - N_u} \qquad (C1)$$

, where $V_{sh,n}$ is the shock propagation speed along the normal direction, $N$ is the plasma number density and $\vec{V}$ is the ion bulk flow velocity.

We determined the Alfvén speed ($v_A$) and the phase speed of the slow magnetosonic wave ($v_{slow}$) using the following equations:

$$v_A = \frac{B_{total}}{\sqrt{\mu_0 N m_p}} \qquad (C2)$$

, where $B_{total}$ is the magnetic field intensity and $m_p$ is the proton mass.

$$v_{slow} = \sqrt{\frac{1}{2}\left(c_s^2 + v_A^2 - \sqrt{(c_s^2 + v_A^2)^2 - 4c_s^2 v_A^2 \cos^2\theta}\right)} \qquad (C3)$$

, where $c_s$ is the sound speed and $\theta$ is the angle between the shock normal and the magnetic field vector. The sound speed was calculated from

$$c_s = \sqrt{\frac{\gamma_\parallel k_B T_\parallel + \gamma_\perp k_B T_\perp}{m_p}} \qquad (C4)$$

, where the adiabatic indices $\gamma_\parallel = 1$ and $\gamma_\perp = 2$, $k_B$ is the Boltzmann constant, and $T_\parallel$ and $T_\perp$ are the parallel and perpendicular temperature of plasma, including both electrons and protons.

The Alfvénic Mach number and Slow Magnetosonic Mach number are the plasma flow speeds downstream and upstream of the shock under shock frame divided by the local Alfvén speed and the slow magnetosonic speed.

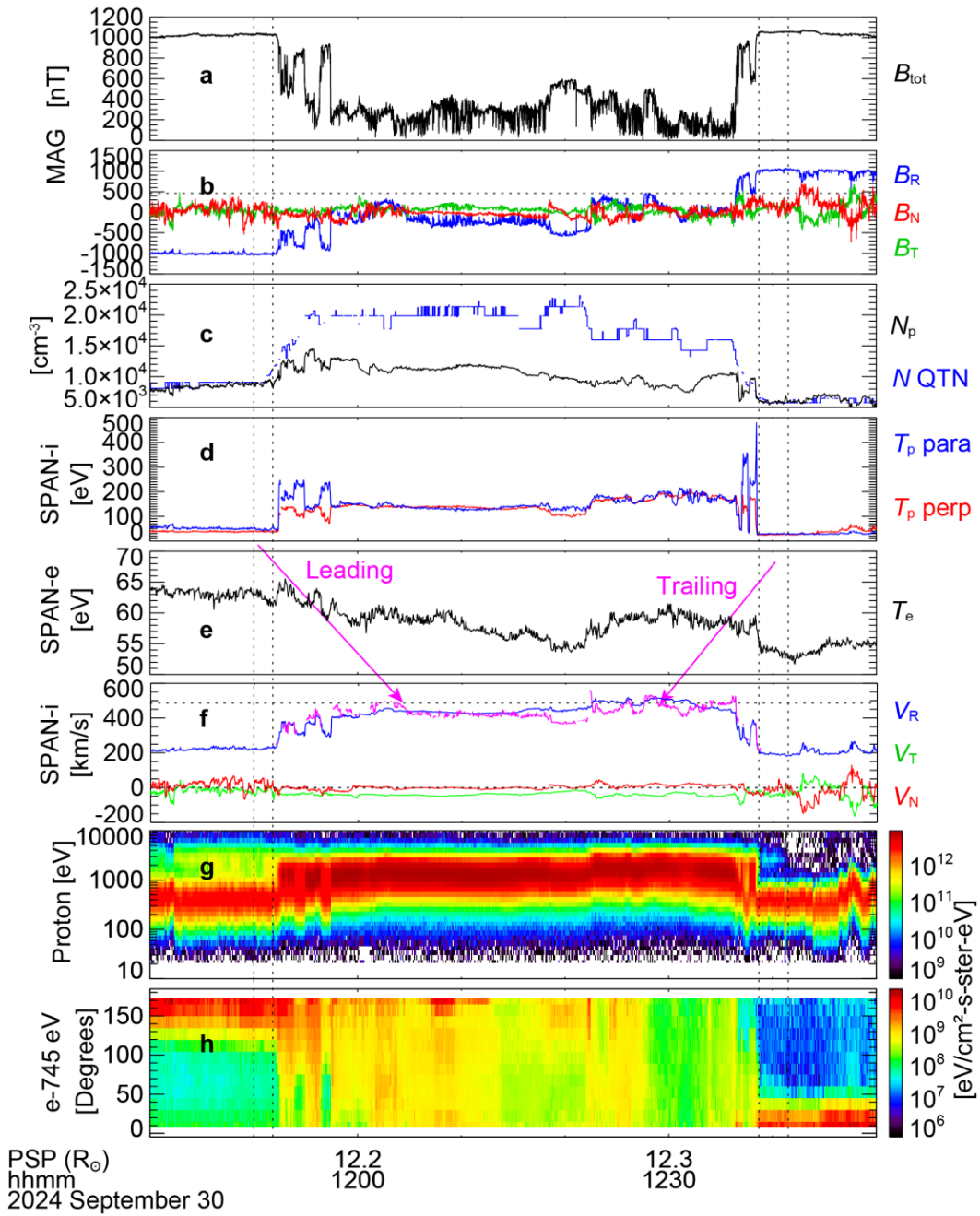

**Figure 1.** Overview of a reconnection outflow in the heliospheric current sheet (HCS) at a distance from 12.2 to 12.3 solar radius ($R_\odot$) that was observed by Parker Solar Probe (PSP) from ~ 11:40 to 12:50, 30 September 2024 UTC. (a) Magnetic field intensity. (b) Magnetic field components. (c) Proton number density from SPAN-ion (black) and electron number density from QTN (blue). (d) Ion temperature. (e) Electron temperature. (f) Ion bulk velocity, magenta line corresponds to the R component of the Walén relations from the external solar wind parameters on the left side (11:50:00 to 11:51:50 UTC) and right side (12:38:40 to 12:41:30 UTC). (g) Proton energy spectrum. (h) Electron pitch angle distribution within the energy channel of 745 eV.

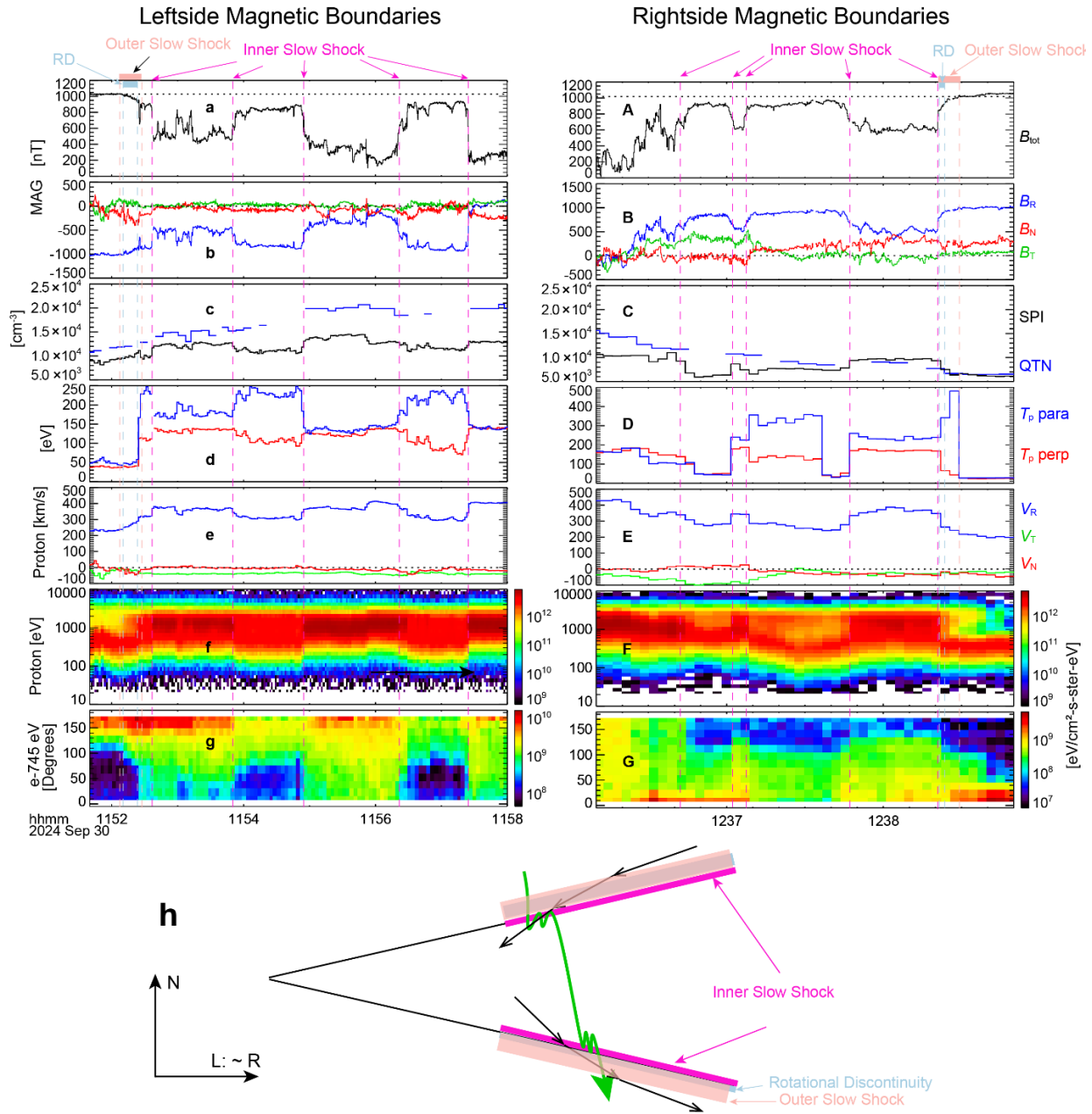

**Figure 2.** Magnetic field and plasma measurements across the magnetic boundaries of the reconnection outflow. (a, A) Magnetic field intensity. (b, B) Magnetic field components. (c, C) Densities from SPAN-ion (black) and QTN (blue). (d, D) Ion temperature. (e, E) Ion bulk velocity components. (f, F) Span-ion energy spectra. (g, G) Span-electron pitch angle distribution of the electron energy channel of 745 eV. (h) A schematic figure of the magnetic boundaries and PSP's trajectory under the local coordinates of the HCS, in which $\vec{L}$ = [-0.987, -0.027, -0.158], $\vec{M}$ = [-0.150, 0.496, 0.855], $\vec{N}$ = [0.055, 0.868, -0.494]. $\vec{N}$ is in the unit vector of the cross product

of the two magnetic field vectors outside the HCS. $\vec{L}$ is the unit vector along the difference between the two field vectors, and $\vec{M}$ completes the right-handed system. We note that the normal directions of the slow shocks could contain large components out of the L-N plane.

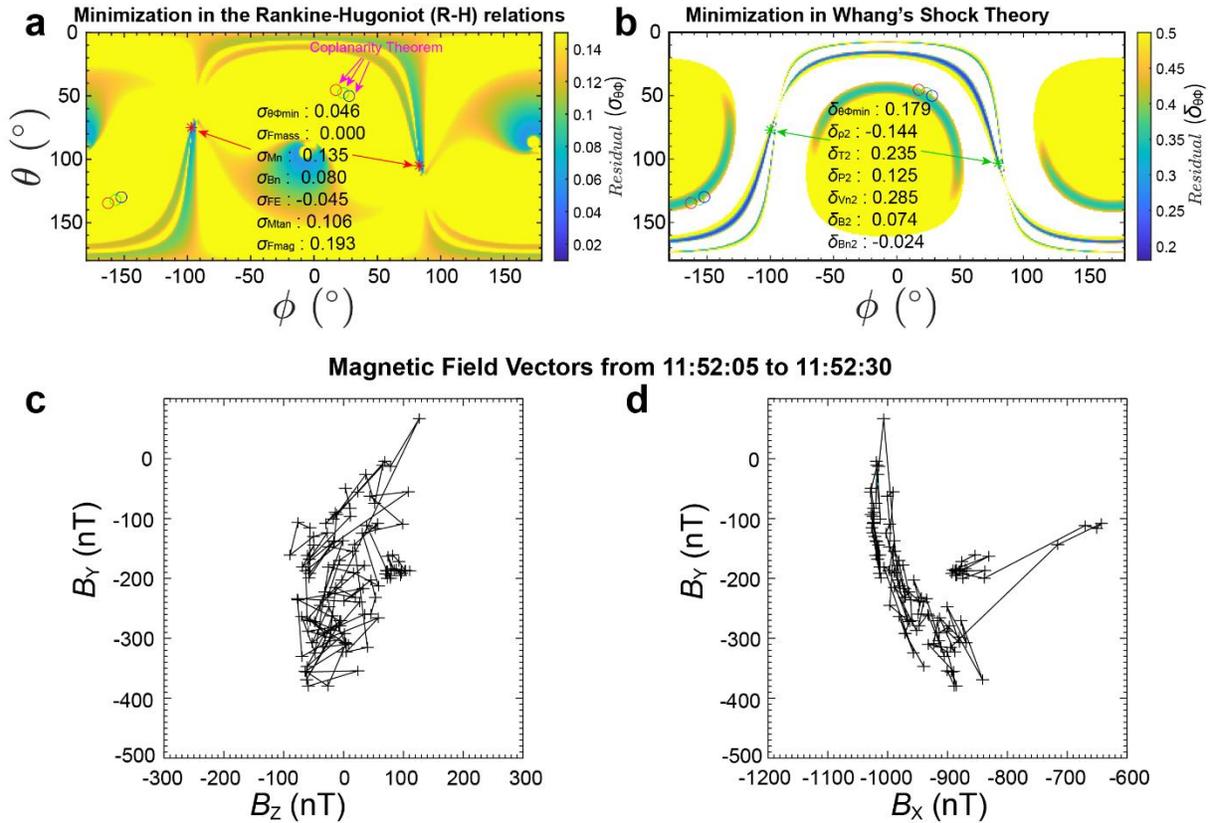

**Figure 3.** Minimization in the Rankine-Hugoniot (R-H) relations (a) and minimization in Whang's shock theory (b). The colors indicate the residual within each 1° × 1° angular grid. The minimum angle is determined to be normal direction of the boundary and are marked in the panels. The directions determined from the coplanarity theorem (Chao 1970; Colburn & Sonett 1966) have been shown in the panels as well. See Appendices for the minimization details.(c, d) Hodograms of the magnetic field component in the $B_Y$-$B_Z$ plane and $B_Y$-$B_X$ plane, respectively. Here $\vec{Z}$=[-0.119, -0.971, 0.208], which is the normal of the outer slow shock, $\vec{Y}$ = [0, 0.209, 0.978], $\vec{X}$ = [0.993, -0.116, 0.025]. The time interval for the hodograms is marked by the blue bar in Figure 2a.

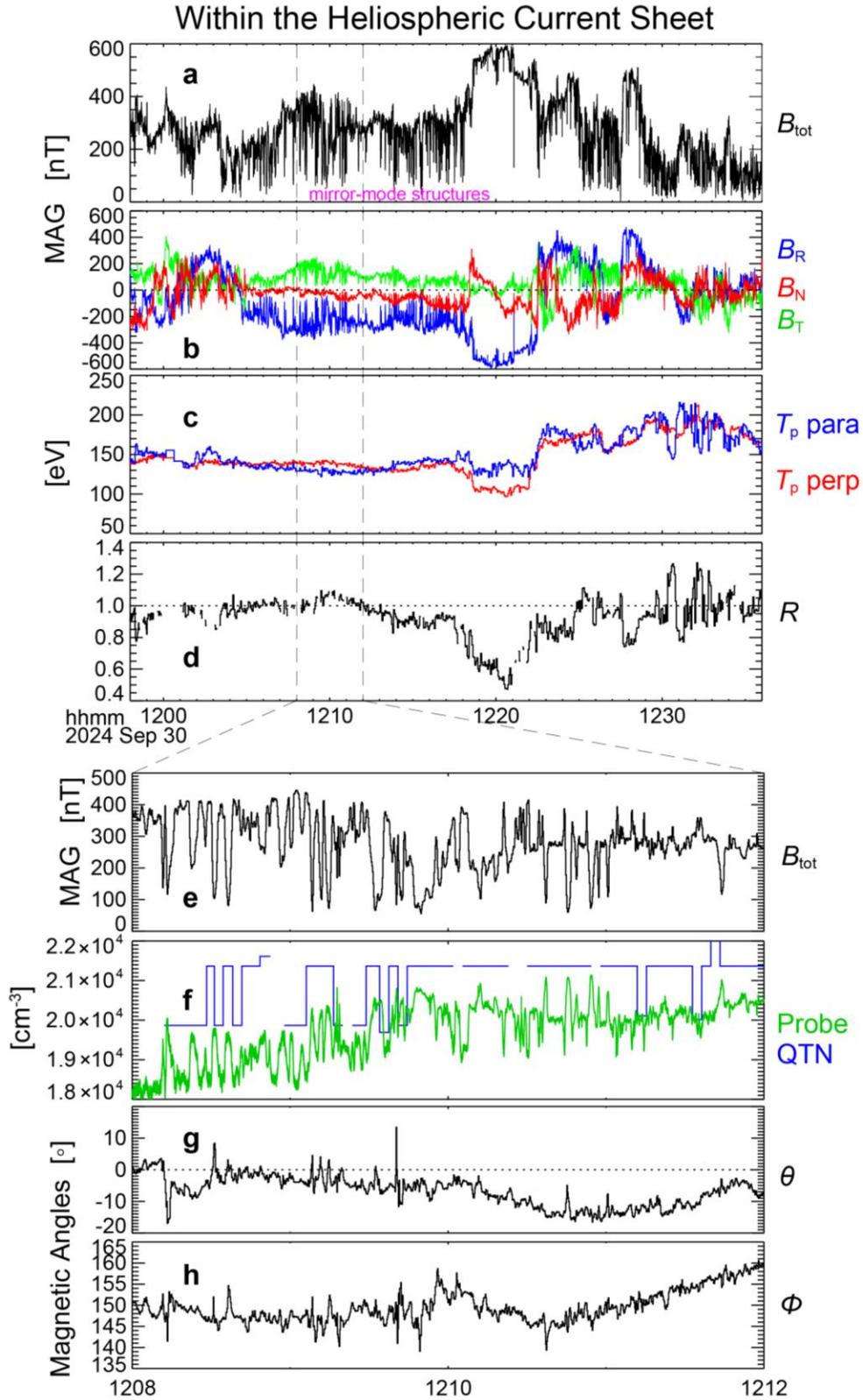

**Figure 4**. Magnetic field and plasma measurements within the HCS of the reconnection outflow. (a, e) Magnetic field intensity. (b) Magnetic field components. (c) Ion temperature. (d) Mirror

instability threshold $R = (T_{perp}/T_{para})/(1 + 1/\beta_{perp})$. (f) Plasma density from probe (green) and QTN (blue). The correlation coefficient between the density and magnetic field intensity is -0.782. (g) magnetic $\theta$ angle ($-90°$ to $90°$). (h) magnetic $\phi$ angle ($-180°$ to $180°$).

Table 1.

| | Time | Results from the Minimization of R-H Relations | | | $M_{SMS}$[a] | $M_A$ | $\theta\langle\vec{n},\vec{B}\rangle$ | Whang's theory |
| | | Normal ($\vec{n}$) | | | | | | Normal ($\vec{n}$) |
| --- | --- | --- | --- | --- | --- | --- | --- | --- |
| LO | 11:52:10 to 11:52:30 | [-0.119, -0.971, 0.208] | | Upstream | 1.09 | 0.57 | 85.7° | [-0.188, -0.970, 0.156] |
| | | | | Downstream | 0.67 | 0.47 | 84.7° | |
| LI | 11:52:35 to 11:52:39 | [-0.215, -0.661, 0.719] | | Upstream | 1.39 | 0.97 | 85.1° | [-0.201, -0.751, 0.629] |
| | | | | Downstream | 0.96 | 0.84 | 81.0° | |
| | 11:53:50 to 11:54:10 | [-0.073, -0.597, 0.799] | | Upstream | 1.08 | 0.80 | 88.1° | [-0.071, -0.448, 0.891] |
| | | | | Downstream | 0.79 | 0.71 | 86.3° | |
| | 11:54:50 to 11:55:00 | [0.017, 0.990, -0.139] | | Upstream | 1.21 | 0.89 | 88.4° | [-0.029, 0.838, -0.545] |
| | | | | Downstream | 0.88 | 0.84 | 86.2° | |
| | 11:56:18 to 11:56:30 | [0.082, -0.462, -0.883] | | Upstream | 1.35 | 0.97 | 88.9° | [0.198, 0.064, -0.978] |
| | | | | Downstream | 0.73 | 0.72 | 85.8° | |
| | 11:57:20 to 11:57:25 | [-0.047, -0.668, -0.743] | | Upstream | 1.35 | 0.98 | 86.1° | [-0.057, -0.654, -0.755] |
| | | | | Downstream | 0.80 | 0.79 | 68.9° | |
| TI | 12:35:40 to 12:36:48 | [0.020, 0.049, 0.999] | | Upstream | 1.80 | 0.97 | 89.0° | [0.020, 0.048, 0.999] |
| | | | | Downstream | 0.57 | 0.57 | 77.8° | |
| | 12:37:00 to 12:37:03 | [0.188, -0.030, -0.982] | | Upstream | 1.52 | 0.73 | 79.6° | [-0.290, 0.181, 0.940] |
| | | | | Downstream | 0.60 | 0.51 | 73.2° | |
| | 12:37:06 to 12:37:10 | [0.407, -0.264, -0.875] | | Upstream | 1.42 | 0.89 | 76.6° | [-0.467, 0.218, 0.857] |
| | | | | Downstream | 0.78 | 0.67 | 68.3° | |
| | 12:37:40 to 12:37:55 | [-0.163, -0.608, -0.777] | | Upstream | 1.39 | 0.81 | 72.0° | [0.182, 0.559, 0.809] |
| | | | | Downstream | 0.99 | 0.81 | 60.8° | |
| | 12:38:20 to 12:38:21 | [0.256, -0.604, 0.755] | | Upstream | 1.34 | 0.78 | 60.7° | [0.050, 0.237, 0.970] |
| | | | | Downstream | 0.83 | 0.71 | 47.9° | |
| TO | 12:38:22 to 12:38:50 | [0.266, -0.730, -0.629] | | Upstream | 1.23 | 0.52 | 88.3° | [-0.266, 0.730, 0.629] |
| | | | | Downstream | 0.86 | 0.48 | 88.0° | |

[a], the calculations of the shock parameters were based on the normal from the minimization of the R-H relations. The shock parameters calculated based on Whang's shock theory (not shown) are consistent with slow shocks for all the boundaries.

4   LO: Leading Outer Boundary; LI: Leading Inner Boundary; TI: Trailing Inner Boundary; TO: Trailing Outer Boundary.
5